\documentclass[conference]{IEEEtran}
\usepackage[utf8]{inputenc}
\usepackage{verbatim}
\usepackage{cprotect}
\usepackage{tikz}
\usepackage{amsmath}
\usepackage{amssymb}
\usepackage{graphicx}

\makeatletter

\providecolor{lyxadded}{rgb}{0,0,1}
\providecolor{lyxdeleted}{rgb}{1,0,0}
\usetikzlibrary{calc}
\newcommand{\lyxobjectsout}[1]{%
  \bgroup%
  \color{lyxdeleted}%
  \tikz{
    \node[inner sep=0pt,outer sep=0pt](lyxdelobj){#1};
    \draw($(lyxdelobj.south west)+(2em,.5em)$)--($(lyxdelobj.north east)-(2em,.5em)$);
  }
  \egroup%
}

\DeclareRobustCommand{\lyxdisplayobjdeleted}[4][]{%
  \ifx#4\empty\else%
     \leavevmode\\%
     \lyxobjectsout{\parbox{\linewidth}{#4}}%
  \fi%
}
\DeclareRobustCommand{\lyxudisplayobjdeleted}[4][]{%
  \ifx#4\empty\else%
     \leavevmode\\%
     \raisebox{-\belowdisplayshortskip}{%
                \lyxobjectsout{\parbox[b]{\linewidth}{#4}}}%
     \leavevmode\\%
  \fi%
}

\IEEEoverridecommandlockouts

\usepackage{cite}
\usepackage{amsfonts}\usepackage{algorithmic}
\usepackage{textcomp}
\usepackage{xcolor}

\usepackage{multicol}

\graphicspath{ {./figures/} }
\def\BibTeX{{\rm B\kern-.05em{\sc i\kern-.025em b}\kern-.08em
    T\kern-.1667em\lower.7ex\hbox{E}\kern-.125emX}}

\makeatother

\begin{document}
\title{An Implementation of Multi-User MIMO Downlink for O-RAN 5G New Radio
using OpenAirInterface}
\author{\IEEEauthorblockN{Duc Tung Bui and Le-Nam Tran} \IEEEauthorblockA{School of Electrical and Electronic Engineering, University College
Dublin, Ireland\\
 Email: duc.t.bui@ucdconnect.ie; nam.tran@ucd.ie}}
\maketitle
\begin{abstract}
We present the first implementation of a Multi-User Multiple-Input
Multiple-Output (MU-MIMO) transmission scheme on the Physical Downlink
Shared Channel (PDSCH) for 5G Open Radio Access Network (O-RAN) based
on OpenAirInterface (OAI). Our implementation features a fully functional
O-RAN-compliant 5G New Radio (5G NR) system, including a 5G Core Network
(5G CN), a refined 5G RAN, which is split into a Centre Unit (CU)
and an Distributed Unit (DU), and 5G NR User Equipment (UEs). This
implementation demonstrates MU-MIMO performance in the downlink while
showcasing the disaggregation capabilities of O-RAN. Specifically,
the Base Station (i.e. gNB) in our setup is capable of serving two
UEs simultaneously over the same downlink Resource Block (RBs). User
scheduling is performed based on the Precoding Matrix Indicators (PMIs)
reported by the UEs according to the NR Channel State Information
(CSI) reporting procedure. The system throughput performance is evaluated
using \textit{iperf}. The obtained results via simulation and testbed
experiments demonstrate that the MU-MIMO scheme achieves significant
downlink throughput gains, particularly in the high Signal-to-Noise-Ratio
(SNR) regime, while keeping the Block Error Rate (BLER) below the
required threshold of $10^{-1}$ for both UEs.
\end{abstract}

\begin{IEEEkeywords}
MU-MIMO, O-RAN, OpenAirInterface, 5G NR, PDSCH.
\end{IEEEkeywords}

\section{Introduction}

Radio Access Networks (RANs) are a critical component of cellular
networks, which connects user mobile devices to the Core Network (CN),
and thus, allows communication between devices and access to the Internet.
RANs play a key role in wireless communication and has undergone significant
evolution in the past two decades, resulting in increased complexity. Subsequently,
the management and optimization of RANs has become more challenging
than ever. However, traditional RANs are typically provided as all-in-one
solutions by a limited number of vendors, creating a \textquotedblleft black
box\textquotedblright{} scenario for operators. This approach limits
operators by tying them to specific vendors, prevents them from re-configuring
and optimizing their networks. It also reduces their flexibility to
deploy and interface RAN equipment from different vendors \cite{RN29}.

To overcome these limitations, in recent years, several research and
standardization efforts have focused on advancing the the O-RAN initiative,
aiming to introduce new levels of openness in next-generation cellular
networks. O-RAN deployments are based on disaggregated, virtualized
and software-based components, connected through open and well-defined
interfaces that support interoperability across different vendors
\cite{RN29,RN26}. Disaggregation and virtualization enable flexible,
cloud-native deployments that improve RAN resiliency and reconfigurability.
Open and interoperable interfaces also allow operators to integrate
equipment from diverse vendors, expanding the RAN ecosystem to include
smaller players. Additionally and more importantly, these open interfaces
and software-defined protocol stacks accelerate the integration of
intelligent, data-driven, closed-loop control mechanisms within the
RAN.

Except for the Radio Frequency (RF) chains, O-RAN is able to be realized
by open-source software stacks which can run on general-purpose computers.
Currently, several open-source projects such as OpenAirInterface5G
(OAI 5G RAN) \cite{RN197,RN196}, srsRAN \cite{RN199,RN187}, and
Open AI Cellular \cite{RN200,RN213} (an expansion of the srsRAN),
are widely used by researchers for 5G O-RAN simulation and deployment.
For this work, we opt for the OpenAirInterface5G RAN because of its
extensive functionalities, as well as its seamless integration with
the OpenAirInterface Core Network (OAI 5G CN) \cite{RN201}, allowing
for the creation of a fully functioning 5G network. In the current
OAI's implementation, a gNB is able to serve multiple UEs using proportional
fair scheduling. Thus, at a given time slot, a gNB can only serve
a single UE with up to $2\times2$ Multiple-Input Multiple-Output
(MIMO) configuration. In other words, the OAI software currently fully
supports up to 2 layers transmission using 2 transmit (TX) antennas
at the gNB and 2 receive (RX) antennas for each UE.

In a mobile communication system with MU-MIMO functionality on the
downlink, a base station with multiple antennas can simultaneously
transmit data to multiple users on the same frequency, and thus, could
provide significant gains over traditional multiple-access systems
\cite{RN85}. Although MU-MIMO has been a focus of considerable interest
\cite{RN210,RN99,RN116,RN171}, there have been a few attempts to
implement this functionality in an open-source 5G framework. For the
downlink, the primary challenge lies in the complexity to compute
the precoding matrices for each UE based on their channel conditions,
to reduce the interference between UEs \cite{RN99}. In \cite{RN215},
the authors demonstrated a MU-MIMO system based on OAI and focused
mainly on analyzing the power and energy consumption of the system.
The MU-MIMO uplink data rate was studied in \cite{RN215} and \cite{RN172}.

In this work, we develop the MU-MIMO functionality on the PDSCH within
the existing OAI 5G RAN implementation. Capitalizing on the flexibility
of the OAI framework, we incorporate the 3GPP-defined Type I Codebook
\cite{RN171}, and revamp the scheduling algorithm to enable MU-MIMO
transmissions based on the PMIs reported by the UEs \cite{RN171}.
To evaluate the performance of this newly developed feature, we build
a fully functional 5G NR network comprising the OAI 5G CN, OAI 5G
RAN, and two UEs. To showcase the disaggregation capability of O-RAN,
the RAN unit is split into a CU and a DU. The DU gNB is equipped with
a Radio Unit (RU) equipped with 2 TX antennas, while each UE has 1
RX antenna, forming a $2\times2$ MU-MIMO system. Our testings, conducted
via the OAI's built-in rfsimulator, show that our newly developed
MU-MIMO system significantly improves downlink performance compared
to the default proportional fair scheduler. This work lays a foundation
for further development of advanced MIMO features within the OAI,
with the potential for significant contributions to the evolution
of O-RAN.

The rest of this paper is organized as follows. In Section \ref{proposed_mumimo}
we describe our proposed scheduling implementation for the MU-MIMO
scheme. Section \ref{impl_sim} details our virtual system configuration
for simulation along with the corresponding results. In Sections \ref{sec:Practical-Implementation},
we present and discuss the expertimental results of the downlink BLERs
and data rates in practical implementation using Software-Defined
Radio (SDR) devices. Finally, Section \ref{sec:Conclusion-and-Future}
concludes the paper.

\section{A Brief Introduction to MU-MIMO for 5G NR Downlink\label{proposed_mumimo}}

\begin{figure}
\centerline{\includegraphics[width=1\linewidth]{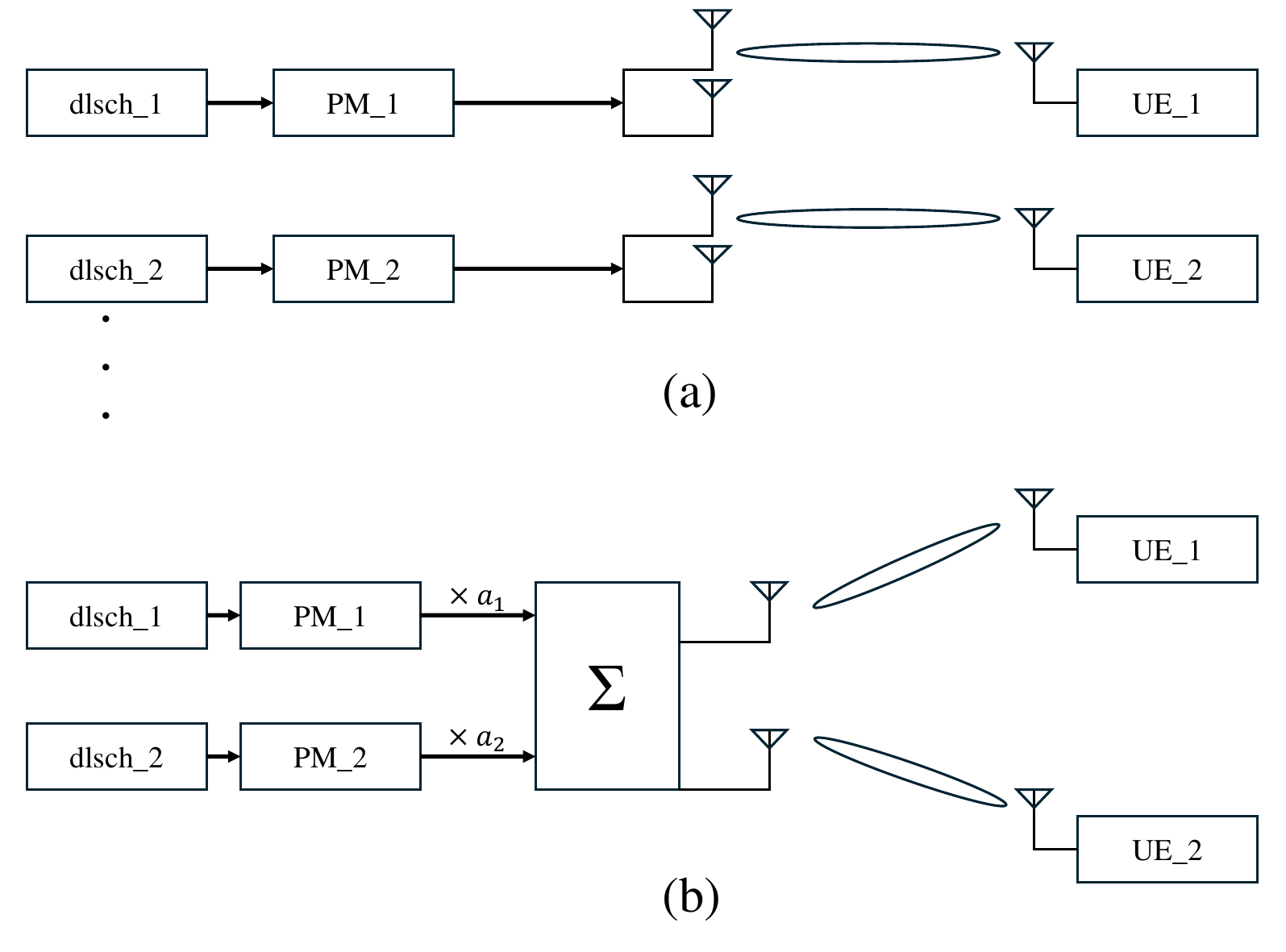}}

\caption{Different MIMO use cases. (a) Single-User MIMO. (b) Multi-User MIMO.}

\label{mu-mimo}
\end{figure}

In this section, we provide a brief overview of MU-MIMO operation
in the 5G NR downlink. As a case study, we consider a MU-MIMO downlink
system consisting of a gNB and two UEs, referred to as \textit{\emph{UE\_1}}
and \textit{\emph{UE\_2}}. The gNB is equipped with 2 TX antennas,
while each UE has a single RX antenna. Hence, the downlink stream
for each UE always has only one layer. We note that the current OAI
5G RAN implementation uses Type I Codebook \cite{RN171} for precoding,
which is simpler than the Type II Codebook \cite{RN171}. However,
this simplification imposes certain constraints on the wireless channels
of the UEs: for MU-MIMO transmission to be effective, the channels
must satisfy specific conditions to allow spatial separation and minimize
inter-user interference.

\begin{figure}
\centerline{`\includegraphics[width=1\columnwidth]{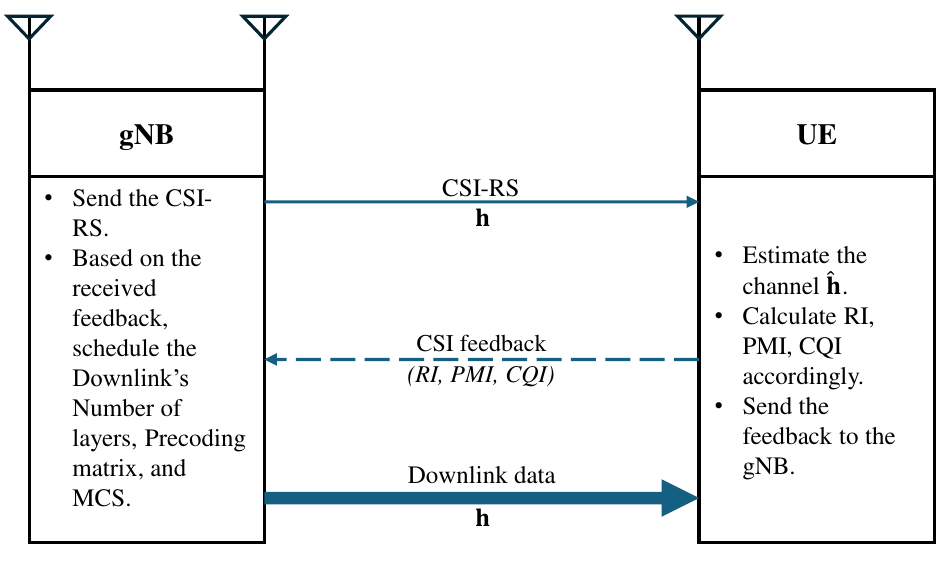}}

\caption{A brief description of the CSI reporting procedure.}

\label{csi_rs}
\end{figure}

The gNB acquires information about the wireless channels via the CSI
reporting procedure \cite{RN208,RN171}. As part of this process,
the gNB first broadcasts a known reference signal---specifically,
the Channel State Information Reference Signal (CSI-RS)\cite{RN208}---
to its UEs. Each UE uses the received CSI-RS to estimate the channel
and then report the CSI parameters back to the gNB \cite{RN171}.
In our system implementation, we adopt the channel estimation algorithm
available within the OAI 5G RAN source code. The resulting CSI feedback
includes the Channel Quality Indicator (CQI), the Rank Indicator (RI),
and the Precoding Matrix Indicator (PMI). Fig. \ref{csi_rs} illustrates
the CSI reporting process between the gNB and a UE, where $\mathbf{h}$
denotes the wireless downlink channel between the two transceivers.
Among the three reported CSI parameters, the PMI is particularly critical,
as it identifies the most suitable precoding matrix from the codebook
for each UE. By comparing the PMI reported by different UEs, the gNB
can decide if MU-MIMO transmission is beneficial.

The Type I Codebook for single-layer transmission with two transmit
antennas is shown in the Table 5.2.2.2.1-1 of \cite{RN171}. From
this table, it can be seen that the precoding matrices corresponding
to indices 0 and 2, as well as those corresponding to indices 1 and
3 are orthogonal. This orthogonality implies that, when the two UEs
report PMIs of 0 and 2, or 1 and 3, respectively, the gNB can schedule
MU-MIMO transmissions for these UEs with minimal inter-user interference.
Such scheduling enables more efficient spatial multiplexing and leads
to improved downlink throughput for the network.

Using the available source code of the OAI 5G RAN, we implement the
MU-MIMO functionality for the 5G NR PDSCH by modifying the scheduling
algorithm. Once the gNB determines that MU-MIMO transmission is beneficial,
it assigns the signal intended for the \textit{\emph{UE\_2}} to the
same set of subcarriers already allocated to \textit{\emph{UE\_1}}.
The symbols for both UEs are then precoding using the respective precoding
matrices and scaled by power coefficients $\alpha_{1}$ and $\alpha_{2}$,
respectively. The resulting precoded and scaled symbols are then summed
together before being transmitted through the gNB\textquoteright s
antennas. Fig. \ref{mu-mimo} summarizes this process and highlights
the primary differences between MU-MIMO and traditional Single-User
 scheduling.

The transmitted signal in the considered MU-MIMO downlink system is
written as:
\begin{equation}
\mathbf{s}=\alpha_{1}\mathbf{w}_{1}x_{1}+\alpha_{2}\mathbf{w}_{2}x_{2}\label{sending_sig}
\end{equation}
where $x_{1}$ and $x_{2}$ are the symbols sent to the \textit{\emph{UE\_1}}
and \textit{\emph{UE\_2}}, respectively, and $\mathbf{w}_{1}$ and
$\mathbf{w}_{2}$ are their corresponding precoding vectors.

In a perfect scenario--- for example, when the wireless channels
of the two UEs are $\mathbf{h}_{1}=\left[\begin{array}{cc}
1 & j\end{array}\right]$ and $\mathbf{h}_{2}=\left[\begin{array}{cc}
1 & -j\end{array}\right]$, the UEs should report PMIs of 3 and 1, respectively. The associated
precoding vectors from Table 5.2.2.2.1-1 of \cite{RN171} are
\begin{equation}
\mathbf{w}^{1}=\frac{1}{\sqrt{2}}\left[\begin{array}{c}
1\\
j
\end{array}\right],\mathbf{w}^{3}=\frac{1}{\sqrt{2}}\left[\begin{array}{c}
1\\
-j
\end{array}\right].
\end{equation}
As a result, the received signals at \textit{\emph{UE\_}}1 and \textit{\emph{UE\_}}2,
denoted by $y_{1}$ and $y_{2}$, are expressed as:
\begin{multline}
y_{1}=\mathbf{h}_{1}\mathbf{s}+n_{1}=\mathbf{h}_{1}(\alpha_{1}\mathbf{w}^{3}x_{1}+\alpha_{2}\mathbf{w}^{1}x_{2})+n_{1}\\
=\alpha_{1}\mathbf{h}_{1}\mathbf{w}^{3}x_{1}+\alpha_{2}\mathbf{h}_{1}\mathbf{w}^{1}x_{2}+n_{1}=\sqrt{2}\alpha_{1}x_{1}+n_{1}\label{recei_sig0}
\end{multline}
\begin{multline}
y_{2}=\mathbf{h}_{2}\mathbf{s}+n_{2}=\mathbf{h}_{2}(\alpha_{1}\mathbf{w}^{3}x_{1}+\alpha_{2}\mathbf{w}^{1}x_{2})+n_{2}\\
=\alpha_{1}\mathbf{h}_{2}\mathbf{w}^{3}x_{1}+\alpha_{2}\mathbf{h}_{2}\mathbf{w}^{1}x_{2}+n_{2}=\sqrt{2}\alpha_{2}x_{2}+n_{2}\label{recei_sig1}
\end{multline}
where $n_{1}$ and $n_{2}$ are AWGN noise at the respective UEs.
To maintain the power per resource block (RB), the power coefficients
are constrained such that $\alpha_{1}^{2}+\alpha_{2}^{2}=1$. In this
work, we consider equal power allocation between the two UEs, and
thus $\alpha_{1}=\alpha_{2}=\sqrt{1/2}$.

Under ideal channel conditions, the MU-MIMO scheme is expected to
perform well, as each UE's receive signal is only affected by its
own channel and white noise. In more realistic situations, some level
of interference between the two UEs would certainly occur. However,
with proper precoding, the effective channels can be made nearly orthogonal,
and thus, the residual interference is small enough for UEs' detectors
to effectively handle.

In our OAI 5G RAN implementation, MU-MIMO scheduling is activated
when the reported PMIs of the UEs are either $\{0,2\}$ or $\{1,3\}$.
The MU-MIMO scheme is scheduled by aligning the starting subcarrier
of the second UE (\textit{\emph{UE\_2}}) with that of the first UE
(\textit{\emph{UE\_1}}). Additionally, it is important to mention
that, once the MU-MIMO scheme is scheduled, the Transport Block size
(TBS) and the Modulation and Coding Scheme (MCS) \cite{RN171} of
the UE\_2 need to be set equal to those of the UE\_1. This ensures
that the same subcarriers are used by both users. To maintain system
stability, MU-MIMO is scheduled only for UE\_2\textquoteright s initial
transmission attempt. If the transmission is unsuccessful and a re-transmission
is required, the system reverts to the default OAI scheduling scheme---namely,
the proportional fair MIMO mode. In the following section, we evaluate
the performance of the proposed implementation.

\section{Virtual deployment and Simulation results\label{impl_sim}}

We first deploy a simulation system using Virtual Machines (VMs),
in order to demonstrate the disaggregation feature of the O-RAN architecture,
as well as to verify the MU-MIMO scheduling algorithm.

\subsection{Deployment\label{subsec:Deployment}}

\begin{figure}
\centerline{\includegraphics[width=1\columnwidth]{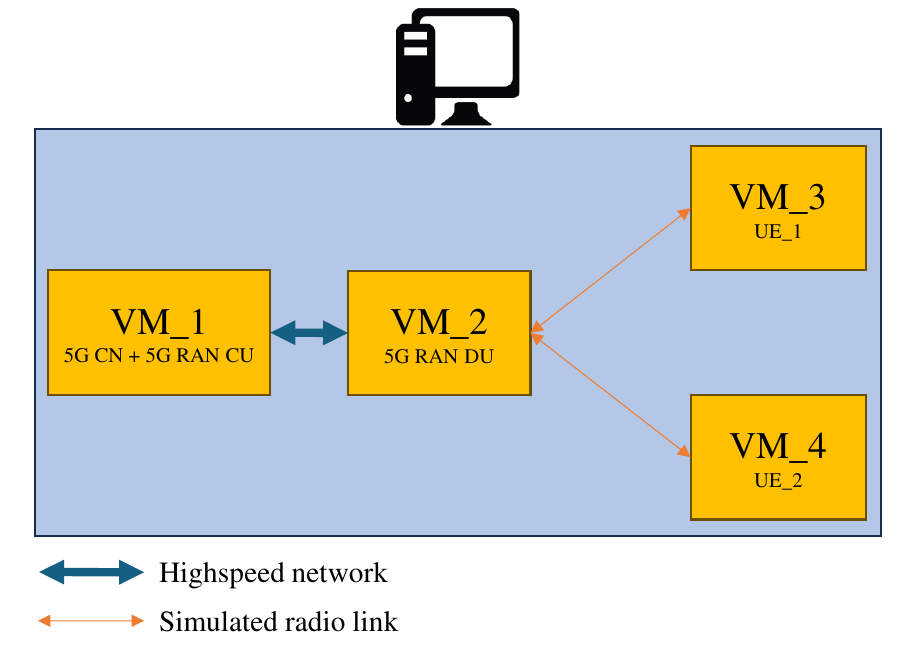}}
\caption{Virtual OAI 5G system configuration}
\label{sys_conf}
\end{figure}

A fully functional OAI 5G NR system includes the 5G CN, 5G RAN, and
5G NR UEs. As the OAI 5G RAN follows the O-RAN architecture, it can
be split into a CU and possibly multiple DUs \cite{RN211}. In practice,
the CU and DUs can be deployed either at the same physical location
or across different sites, as disaggregation is a fundamental principle
of the O-RAN framework.We set up four VMs in our environment. The
first VM hosts the 5G CN and 5G RAN CU, while the second one runs
the 5G RAN DU. These two VMs constitute the gNB, and the remaining
two VMs serve as the UEs. As described in the Section \ref{proposed_mumimo},
the \textit{\emph{gNB}} is equipped with 2 TX antennas, while each
UE has 1 RX antenna. The overall system configuration is illustrated
in Fig. \ref{sys_conf}. The radio links between the \textit{\emph{gNB}}
and the two UEs are simulated using the rfsimulator module available
within the OAI 5G RAN. All the VMs are KVM-based and hosted on a Dell
Alienware R16 equipped with an Intel(R) Core(TM) i9-14900F CPU, 64GB
of RAM, 512GB SSD storage, and running on Ubuntu 24.04 x86 64 bit.

We configure our system to operate on band \textit{n78}, which belongs
to Frequency Range \textit{FR1} and uses the Time Division Duplex
(TDD) mode \cite{RN203}. In terms of bandwidth, the system is set
to utilize 106 resource blocks (RBs), with subcarrier spacing of 30
kHz, corresponding to a bandwidth of 40 MHz, including the guardband.
The 5G NR frame structure is described in detail in \cite{RN208}.

As the wireless channels are simulated, a direct measurement of data
rates from the \textit{iperf} output is inappropriate since the result
also includes the time taken by the channel simulation, which should
be excluded. Instead, we use \textit{iperf} to transfer a fix amount
of data from the 5G CN on \textit{\emph{VM\_1}} down to the UEs on
VM\_3 and VM\_4, and we measure the number of RBs used to complete
this transmission. If the number of RBs per time frame is known, the
actual transmission time can be calculated accordingly. Thus, the
total downlink data rate sent by the gNB is determined based on the
data volume and the calculated transmission time. In this work, we
focus specifically on evaluating downlink throughput performance using
the TCP protocol.

As mentioned in Section \ref{proposed_mumimo}, to ensure system stability,
MU-MIMO scheduling is restricted to specific wireless channel conditions
between gNB and the UEs. Hence, instead of relying on randomly generated
channel models, we employ fixed channel models in this work to clearly
showcase the benefits of MU-MIMO transmission. In addition, because
the MCS selection algorithm for the MU-MIMO downlink transmission
has not been fully developed, data rate measurements are taken at
fixed MCS values. Using MCS table 5.1.3.1-1 in \cite{RN171}, we assess
the MU-MIMO performance across MCS values ranging from 10 to 28, under
varying SNR conditions. MCS values below 10 are excluded from testing,
as they are more suitable for low-SNR scenarios, in which the proportional
fair scheduling scheme is preferred for maintaining stable connections.

\subsection{ Simulation results\label{subsec:Perfect-channel-condition}}

In this subsection we evaluate MU-MIMO scheduling under ideal channel
conditions. To this end, the source code of the OAI 5G RAN is modified
to fix the two channels of the two UEs as follows:
\begin{subequations}
\label{eq:idealchannels}
\begin{align}
\mathbf{h}_{1} & =\left[\begin{array}{cc}
1 & j\end{array}\right]\label{chan_0_per}\\
\mathbf{h}_{2} & =\left[\begin{array}{cc}
1 & -j\end{array}\right].\label{chan_1_per}
\end{align}
\end{subequations}
Additive white Gaussian noise (AWGN) is added to the received signals
at the UEs. The noise power is normalized relatively to the power
of the received CSI-RS. Given that both channels have the same gain,
the received CSI-RS power is expected to be equal at both UEs.

\begin{figure*}
\centerline{\includegraphics[clip,scale=0.9]{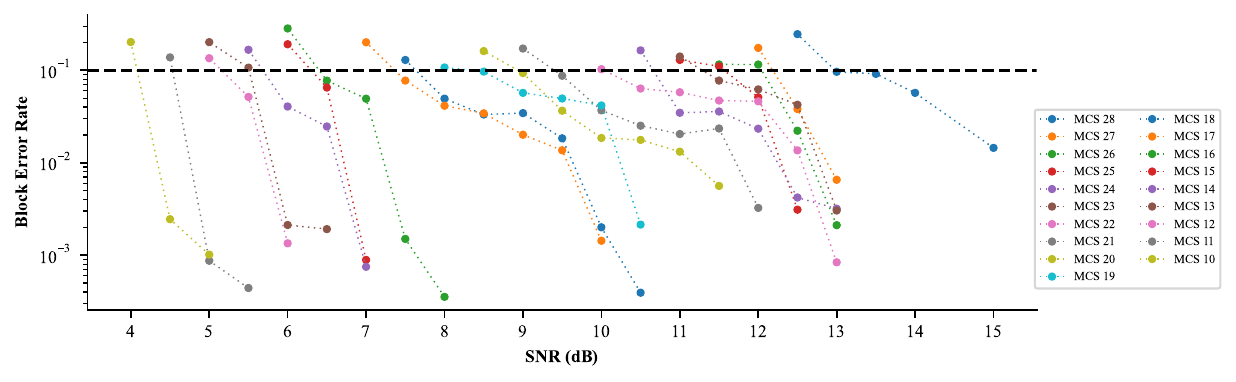}}

\caption{Average downlink BLER of the two UEs versus SNR for the idealized
user channels given in \eqref{eq:idealchannels}.}

\label{bler_perfect}
\end{figure*}

\begin{figure*}
\centerline{\includegraphics[scale=0.9]{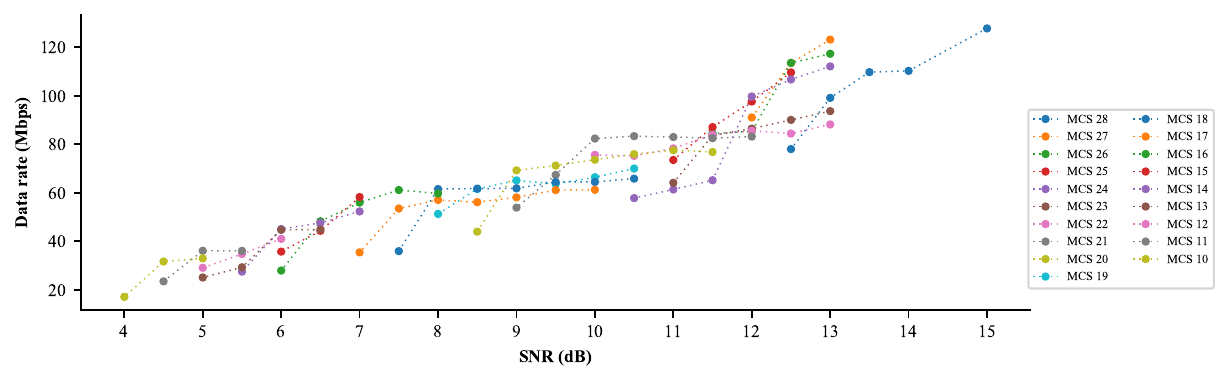}}

\caption{Total downlink data rate versus SNR for the idealized user channels
given in \eqref{eq:idealchannels}.}

\label{datarate_perfect}
\end{figure*}

Figs. \ref{bler_perfect} and \ref{datarate_perfect} shows the average
BLERs%
{} of the two UEs and the total downlink data rate measured at the gNB,
respectively. Note that a transmission is considered reliable when
the BLER is below $10^{-1}$. As shown in Figs. \ref{bler_perfect}
and \ref{datarate_perfect}, for each level of SNR, the MU-MIMO downlink
transmission is able to achieve both reliability and high data rates
through appropriate selection of the MCS.

\begin{figure}
\centerline{\includegraphics{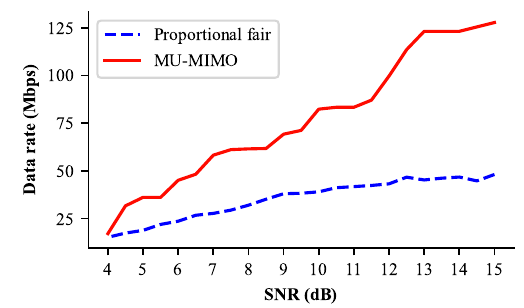}}
\cprotect\caption{%
Comparison of the data rate achieved by MU-MIMO and proportional fair
scheduling for the idealized user channels given in \eqref{eq:idealchannels}.}

\label{compare_gnb_per}
\end{figure}

Fig. \ref{compare_gnb_per} compares the downlink data rate achieved
by the proportional fair transmission scheme with the achievable downlink
data rate of the MU-MIMO scheme for different levels of SNR. The term
``achievable downlink data rate'' refers to the highest data rate
obtained at each value of SNR, as shown in Fig. \ref{datarate_perfect}.
Clearly, the MU-MIMO scheme on the PDSCH offers a considerable enhancement
in data rate, especially at high SNRs.

\section{Practical Implementation\label{sec:Practical-Implementation}}

After verifying the MU-MIMO algorithm by simulation, we implement
the over-the-air transmission using the Universal Software Radio Peripherals
(USRPs) from National Instruments. We utilize three USRPs B210 in
our testbed. One acts as the RU of the gNB, and it is equipped with
2 TX and 2 RX antennas. The gNB is monolithic in this scenario, meaning
that the OAI 5G CN and OAI 5G RAN are hosted on the same computer.
The other two USRP devices, each equipped with one TX and one RX antenna,
are connected to separate laptops and serve as the UEs. We continue
configuring the system to operate on the band \textit{n78,} with subcarrier
spacing of 30 kHz. However, due to the limitation of the USRP B210,
the bandwidth is bounded to 20 MHz, corresponding to 52 RBs. Similar
to the previous section, we focus on evaluating downlink throughput
performance using the TCP protocol.

\begin{figure}
\centerline{\includegraphics{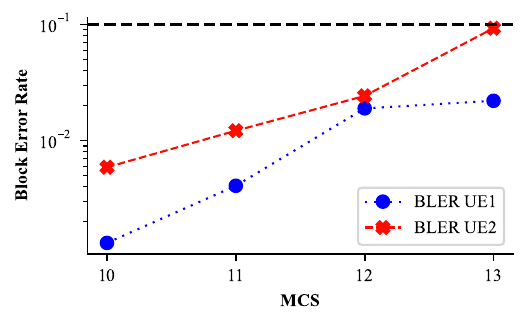}} \caption{Downlink BLER of the two UEs in the practical MU-MIMO mode.}
\label{usrp_bler}
\end{figure}
\begin{figure}
\centerline{\includegraphics{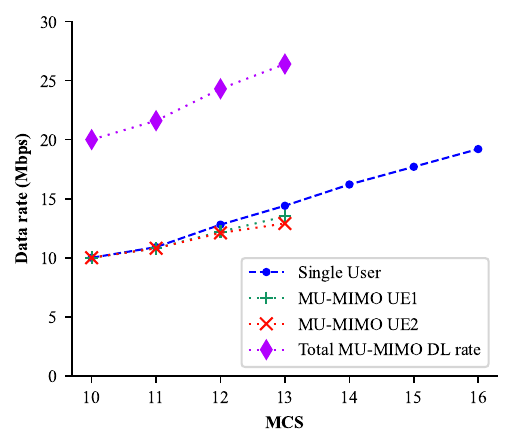}} \caption{Practical downlink data rate using different MCS values.}
\label{usrp_datarate}
\end{figure}

In practice of our system, for single user, the downlink transmission
is only reliable up to the MCS value of 16, before the BLER becomes
too high, and the data rate declines as a result. When the system
is working on the MU-MIMO mode, the maximum downlink MCS for reliable
transmission is 13, as shown in Fig. \ref{usrp_bler}.

In Fig. \ref{usrp_datarate}, the \textit{Single User} line shows
the data rate when the gNB only serves one of the UEs regarding to
each value of MCS. In other words, in this case, one single UE occupies
all the RBs. Hence these rates are equivalent to the maximum total
rates achievable at a particular MCS using Proportional Fair scheduling.
The \textit{MU-MIMO UE1} and \textit{MU-MIMO UE2} lines illustrate
the downlink data rate received by each UE in MU-MIMO transmission
mode. Clearly, both UEs receive the downlink rate nearly equal to
the single user case. Consequently, the total downlink rate, demonstrated
by the line \textit{Total MU-MIMO DL rate}, is improved significantly,
even in comparison with higher MCS values.

\section{Conclusion\label{sec:Conclusion-and-Future}}

In this paper, we have presented our implementation of MU-MIMO functionality
for the 5G PDSCH within the OAI 5G RAN framework. To evaluate its
performance, we have built and tested a fully functional 5G NR system
comprising the OAI 5G CN, a modified OAI 5G RAN, and two OAI 5G UEs.
Our current virtual environment configuration also demonstrates the
disaggregation capability of the O-RAN architecture by splitting the
RAN into a CU and a DU, deployed on separate virtual machines. 

In both simulated and testbed systems, the\textit{\emph{ gNB}} with
2 TX%
{} antennas serves two UEs, each equipped with a single RX antenna,
at the same time, and on the same frequency resources. In other words,the
\textit{\emph{gNB}} can send the downlink data simultaneously to both
UEs over the same RBs by leveraging MU-MIMO. The MU-MIMO downlink
transmission is scheduled if the PMIs reported by the UEs are orthogonal.
Our simulation and experimental results show that, under the considered
scenarios, the MU-MIMO scheme can reliably transmit data, keeping
the downlink BLERs below the acceptable threshold of $10^{-1}$. Moreover,
the MU-MIMO-enabled system significantly improves the total downlink
throughput, especially  in the high-SNR regime where the noise power
is substantially lower than the received signal power. Furthermore,
since the number of RBs remains unchanged when switching between the
Proportional Fair and MU-MIMO schemes, the improvement in throughput
indicates a proportional increase in the spectral efficiency of each
subcarrier.

\section*{Acknowledgment}

This publication has emanated from research supported by Taighde Éireann
- Research Ireland under Grant numbers 22/US/3847 and 13/RC/2077\_P2
at \emph{CONNECT: the Research Ireland Centre for Future Networks}.

\bibliographystyle{IEEEtran}
\bibliography{IEEEabrv,mybib}
 \vspace{12pt}

\end{document}